\documentclass[twocolumn,showpacs,preprintnumbers,amsmath,amssymb]{revtex4}

\usepackage{graphicx}
\usepackage{dcolumn}
\usepackage{bm}

\begin{document}

\preprint{APS/123-QED}

\title{Quantum state measurements using multi-pixel photon detectors}
\author{I. Afek}
\email{itai.afek@weizmann.ac.il}
\author{A. Natan}%
\author{O. Ambar}%
\author{Y. Silberberg}%

\affiliation{%
Department of Physics of Complex Systems, Weizmann Institute of
Science, Rehovot 76100, Israel }
\date{\today}

\begin{abstract}
The characterization and conditional preparation of multi-photon quantum states requires the use of photon number resolving detectors. We study the use of detectors based on multiple avalanche photodiode pixels in this context.
We develop a general model that provides the positive operator value measures for these detectors. The model incorporates the effect of cross-talk between pixels which is unique to these devices. We validate the model by measuring  coherent state photon number distributions and reconstructing them with high precision. Finally, we evaluate the suitability of such detectors for quantum state tomography and entanglement-based quantum state preparation, highlighting the effects of dark counts and cross-talk between pixels.
\end{abstract}

\maketitle

\section{\label{sec:level1}Introduction}
The development of detectors with photon number resolution in the ''few'' photon regime is of key interest in quantum optics. Such detectors play a central role in the preparation and characterization of non-classical multi-photon states. Experiments requiring photon number resolution include the generation of entangled photon sources via spontaneous parametric down conversion (SPDC) with strong pump fields \cite{PRL04}, linear optics quantum computation \cite{Nat01} and a proposal for beating the Heisenberg limit of interferometry with mixed SPDC and coherent state fields \cite{PRL08}.

The realization of photon number resolving detectors is extremely challenging. The benchmark properties for these are quantum efficiency, dark count rate, dead time, operating temperature and the degree and quality of photon number resolution \cite{CP07}. One of the first realizations of a photon number resolving detector was the 'visible light photon counter'  (VLPC) \cite{APL99}, an avalanche photo-diode whose output pulse height is proportional to the number of detected photons due to low multiplication noise in the avalanche process. It has extremely high detection efficiency of around $90\%$ but requires cryogenic cooling and exhibits a high dark rate. More recently, super-conducting devices have been investigated. These devices have high quantum efficiencies and low dark-counts, but they too require cryogenic cooling to reach extremely low working temperatures \cite{APL98,OPL06,OE05,PRA05}. In addition, a few groups have realized a limited degree of photon number resolution using standard detectors at room temperature \cite{RCI04,NAPH08}. Another approach involves the use of independent single photon detectors at the output ports of a balanced array of beam-splitters \cite{PRA01,PRL96}. Although theoretically tractable, this solution is experimentally unappealing due to the large number of detectors involved and the complexity of the setup. A practical way to overcome this difficulty has been the use of temporal modes instead of spatial ones \cite{PRA03,JMODOP04}. One can separate photons into $16$ separate temporal modes using only two detectors.

An alternative method for separating photons into distinct spatial modes is allowing the beam to spread via free space propagation onto an array of single photon detectors. Each detector in the array provides a binary 'click' when one or more photons are detected by it. When the total number of photons is significantly smaller than the number of detectors then it may be assumed that each pixel detected at most one photon. In this regime, the number of pixels that ''clicked'' is equal to the number of detected photons. Clearly, high efficiency avalanche photo-diodes (APDs) are favorable candidates for pixels. Recently, the fabrication of an array of APD pixels for multi-photon detection has been realized \cite{OPEX07,IEEENS07}. These detectors which we refer to as ''multi-pixel detectors'' have been studied extensively for use in particle physics experiments such as the International Linear Collider (ILC) and T2K \cite{NI08}. In this paper we examine, both experimentally and theoretically, the use of multi-pixel detectors from the point of view of the quantum optics community. These detectors are appealing in that they exhibit a high single pixel efficiency, fast detection rates and excellent photon number resolution. Two limitations which must be taken into consideration are the relatively high dark-count rate and the possibility of cross-talk between pixels. We note that an intensified CCD camera has been previously used for photon counting with number resolution \cite{PRA05a}. Such cameras however, have slow repetition rates and a quantum efficiency lower than APDs.

In sections \ref{sec:experimentalsetup} and \ref{sec:dataaquisition} we describe the experimental setup and our method for data acquisition which minimizes the room temperature dark-count rate. In section \ref{sec:theoryanddataanalysis} we develop a theoretical model for relating the measured data to the actual photon number distribution. In section \ref{sec:Coherentstate} we validate the model by measuring coherent state statistics. Finally, in section \ref{sec:quantummetrology}, we evaluate the applicability of multi-pixel detectors to quantum tomography and entanglement-based quantum state preparation.

\section{\label{sec:experimentalsetup}Experimental Setup}
\begin{figure}[hb]
\includegraphics{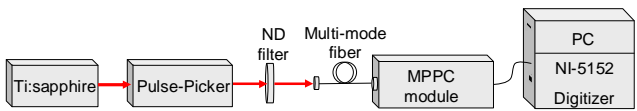}
\caption{\label{fig:setup} Setup for evaluation of MPPC.}
\end{figure}

The Hamamatsu multi-pixel photon counter (MPPC) serves as our prototype detector. We use a module (C10507-11-050U) that incorporates the $400$ pixel MPPC and peripheral electronics. Our experimental setup is shown in Fig. \ref{fig:setup}. The light source is a mode locked Ti-Sapphire oscillator (Spectra Physics, Tsunami) which emits $120$ fs pulses centered at $810$ nm with a $80$ MHz repetition rate. The detector's output pulses are $\sim 20$ ns FWHM, therefore the maximum usable laser rep. rate is $\sim 10$ MHz. We use a pulse picker (PulseSelect, APE-Berlin) to reduce the laser repetition rate to $1$ MHz. After strongly attenuating the beam using ND filters we allow the remaining light to impinge on a multi-mode fiber which is coupled directly to the detector. The detector output signal is connected to a PC-based fast digitizer (NI PCI-5152).

\section{\label{sec:dataaquisition}Data Aquisition}
\begin{figure}[h]
\includegraphics{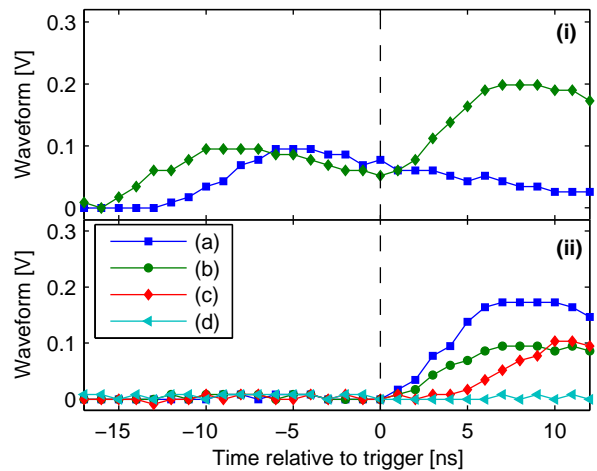}
\caption{\label{fig:digitizer}  Examples of waveforms obtained
from the MPPC. The vertical line marks the arrival time of the
illuminating pulse. (i) Waveforms which are rejected since they are
overrun by dark-counts. (ii) Selected waveforms used in the data-analysis: (a) A two avalanche
waveform; (b) A one avalanche waveform; (c) A zero avalanche
waveform which starts to rise outside the designated time-window due to a dark-count;
(d) A zero avalanche waveform that is flat during the whole time window;}
\end{figure}

The output of a multi-pixel detector is the sum of the outputs of each of the pixels. Thus the height of a given output pulse is proportional to the number of pixels which ''clicked'' in a given detection event.  Multi-pixel detectors typically suffer from a high overall dark count rate. This is because the total dark count rate is a sum of the dark counts of all the pixels. In our experiment the computer based digitizer receives a gating signal from the pulse picker which is synchronized with the pulse arrival times. For each trigger we acquire a $20$ ns waveform at a $1$GHz sampling rate in pre-trigger mode starting from $5$ ns before the trigger arrival time. We use the following scheme to post-select valid waveforms:
\begin{enumerate}
 \item If the waveform is not at the digitizers' zero level $1$ ns before the arrival of the trigger then the signal is discarded. This implies that a dark-count occurred prior to the trigger and may interfere with the readout process by obscuring the signal from the actual pulse or causing an undesirable after-pulse (see Fig. \ref{fig:digitizer}(i)). Otherwise continue to step two.
 \item If a rising edge occurs within $3$ ns of the trigger arrival then the waveform value is saved at some time close to its peak (e.g. $t=5$ ns). This value is later used for determining the number of photons in the pulse. If no rising edge occurs then the value $0$ is saved, corresponding to a zero photon event since no pixels fired in the designated time window (see for example Fig. \ref{fig:digitizer} (ii) curves c and d).
 \end{enumerate}

 This detection scheme ensures the rejection of most of the dark-count and after-pulse events. The obtained single photon  dark count rate is $2.3 \times 10^{-3}$ counts per pulse. The dark count rate without the post-selection scheme described  above is $2.2 \times 10^{-2}$ i.e. almost an order of magnitude higher. We note that dark counts may be significantly reduced using liquid nitrogen cooling \cite{arxiv08}. Using the described method we process data at a rate of $80$ KHz, limited mainly by the digitizers' internal buffer size. Using the largest available buffer available for the NI PCI-5152 digitizer series would allow a much higher data processing rate.

\begin{figure}
\includegraphics{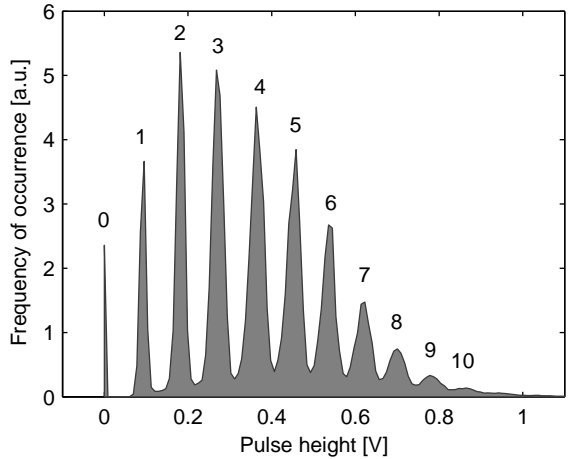}
\caption{\label{fig:pulseheight1} A pulse height histogram obtained
using a coherent state light source. The well distinguished peaks indicate accurate photon number resolution.}
\end{figure}

 Fig. \ref{fig:pulseheight1} shows a histogram of the pulse heights obtained when illuminating with a coherent state. The MPPC is shown to exhibit a high degree of photon number resolution manifested in the well separated peaks. Events of up to $10$ pixels firing simultaneously are easily resolved.

\section{\label{sec:theoryanddataanalysis}Theory and Data-analysis}
The probability $p'_n$ for a given  number
of avalanches per pulse is proportional to the area underneath the
respective peak in the histogram in Fig.
\ref{fig:pulseheight1}. In this section we solve the
problem of finding the relation between $p'_n$ and the actual
photon number distribution impinging on the detector which we denote
$p_n$. This relation is a linear transformation and can be conveniently written
in matrix form \cite{JMODOP04,PRA03}. For multi-pixel detectors, the transformation has three contributions
originating from losses, cross-talk and dark-counts.
\subsection{Loss}
The photon detection efficiency (PDE) of a multi-pixel detector can be obtained by multiplying the PDE of a single pixel by a geometric factor that accounts for the unutilizeduneffective area between the pixels. According to specifications \cite{hamamatsu}, the geometric factor for the $400$ pixel device is $61.5\%$ and the total PDE, $\eta$, has a peak value of $50\%$ at $400$ nm and drops to $8\%$ at $800$ nm. Using the definitions introduced above, we find that loss can be compensated using the following vectorial relation between $\mathbf{p}$ and $\mathbf{p}'$:
\begin{gather}
\mathbf{p}'=\mathbf{M}_L\cdot \mathbf{p}  \notag
\\
[M_L]_{n,m}=\left(
\begin{array}{c}
m\\
n
\end{array}\right)\; \eta^{n}\left( 1 - \eta \right)^{m-n}\,\notag\\
\qquad \qquad \qquad \qquad \qquad \qquad \qquad n,m=0,1,2\dots
\label{eq:loss},
\end{gather}
where $\mathbf{M}_L$ is a square matrix used to describe the effects
of loss.

\subsection{Dark Count}
Dark counts are avalanches in APD pixels that originate from thermal noise. In principle a dark count can consist of multiple avalanches from different pixels. We consider only single avalanche dark counts since the effect of higher order dark counts is negligible. Denoting the probability per pulse for a single avalanche dark count by $\varepsilon_{\mbox{\fontsize{5}{8}\selectfont ${D}$}}$, the relation for
dark-count compensation is
\begin{gather}
\mathbf{p}'=\mathbf{M}_{D}\cdot \mathbf{p} \notag \\
[M_{D}]_{n,m}=\begin{cases}
1-\varepsilon_{\mbox{\fontsize{5}{8}\selectfont ${D}$}} & \text{if m = n;}\\
\varepsilon_{\mbox{\fontsize{5}{8}\selectfont ${D}$}} & \text{if m = n - 1;}\\
0 & \text{otherwise.}
\end{cases}\notag
\\
\qquad \qquad \qquad \qquad \qquad \qquad \qquad n,m=0,1,2\dots
\end{gather}
Note that $\varepsilon_{\mbox{\fontsize{5}{8}\selectfont ${D}$}}$ is the dark-count probability which would occur in the absence of cross-talk (see Eq. (\ref{eq:dark-count})).

\subsection{Cross-talk}
Due to the proximity of pixels, an avalanche in a given pixel may induce avalanches in its neighbors. We use a model with one parameter, $\varepsilon_{\mbox{\fontsize{5}{8}\selectfont ${XT}$}}$, which is the probability that a given avalanche will induce cross-talk. We find the following relation for cross-talk compensation:
\begin{gather}
\mathbf{p}'=\mathbf{M}_{XT}\cdot \mathbf{p}  \notag
\\
[M_{XT}]_{n,m}=\left(
\begin{array}{c}
n\\
n-m
\end{array} \right)\; \varepsilon_{\mbox{\fontsize{5}{8}\selectfont ${XT}$}}^{n-m}\left( 1 - \varepsilon_{\mbox{\fontsize{5}{8}\selectfont ${XT}$}} \right)^{m} \notag \\
\qquad \qquad \qquad \qquad \qquad \qquad \qquad n,m=0,1,2\dots \, .
\label{eq:loss}
\end{gather}
Here, $\mathbf{M}_{XT}$ is the cross-talk matrix. Note the similarity of the matrix elements of $\mathbf{M}_{XT}$ and $\mathbf{M}_L$. The main difference being that $\mathbf{M}_{XT}$ is upper trigonal and  $\mathbf{M}_\textsl{L}$ is lower trigonal implying that loss transfers probability from high photon numbers to lower ones whereas cross-talk does the opposite. The compensation for cross-talk was dealt with previously by \cite{OPEX07,arxiv08} in the form of a recursion relation. Our formalism allows for more than one cross-talk event per-pulse. For example, three photons may initiate up to six avalanches. The matrix for first-order-only cross-talk may be constructed by using the diagonal and sub-diagonal elements of $\mathbf{M}_{XT}$ only.  We show below (Fig. \ref{fig:pulseheight2}) that high order cross-talk events are crucial for fitting high photon numbers whose probabilities are low. In addition, the matrix form of our method allows the relation between $\mathbf{p}$ and $\mathbf{p}'$ to be
conveniently inverted (see Section \ref{sec:photonnumberreconstruction}).

\subsection{\label{sec:photonnumberreconstruction}Photon Number Reconstruction}
There are two approaches to reconstruction of the photon number distribution. The first approach requires assumed knowledge about the photon number distribution of the light illuminating the detector (i.e. coherent state, spontaneous parametric down-conversion etc...). Usually the distribution will have a parameter, such as mean photon number, which must be chosen correctly in order to fit the measured data. The relation between the assumed distribution $\mathbf{p}$ and the measured data $\mathbf{p'}$ is given by application of the loss, dark-count and
cross-talk matrices in the correct order,
 \begin{gather}
\mathbf{p}'=\mathbf{M}_{XT}\cdot\mathbf{M}_{D}\cdot\mathbf{M}_{L}\cdot
\mathbf{p} \, . \label{eq:pt2p}
\end{gather}
Using Eq.(\ref{eq:pt2p}) we verify that the assumed distribution is compatible with the measured data thus allowing us to find the distribution's parameters.

The second approach for retrieving the photon number distribution requires no a-priori assumptions. In this case we use the inverse of the relation in Eq.
(\ref{eq:pt2p}) to obtain $\mathbf{p}$ from $\mathbf{p}'$,
\begin{gather}
\mathbf{p}=\mathbf{M}^{-1}_{L}\cdot\mathbf{M}^{-1}_{D}\cdot\mathbf{M}^{-1}_{XT}\cdot
\mathbf{p}' \, . \label{eq:p2pt}
\end{gather}
 All of the matrices involved can be inverted analytically as shown previously for the case of loss and dark counts \cite{JMODOP04a}. This second approach is less stable than the first, since the inverted relation in Eq. (\ref{eq:p2pt}) is very sensitive to noise in the measured statistics. For photon numbers with small probabilities this can yield un-physical oscillations in the obtained photon number distributions and negative probabilities. The inversion may be stabilized using physical assumptions about the result \cite{PRA06,NAPHYS09}.

\section{\label{sec:Coherentstate}Coherent State Statistics and Determining the Cross-Talk Probability}
It is well known that the state of light emitted by a pulsed laser
is a coherent state with the following photon-number statistics,
\begin{equation}
\displaystyle p_k = e^{ -\langle n \rangle} \frac{ \langle n
\rangle^k}{k!}\, ,
\end{equation}
where $\displaystyle \langle n \rangle$ is the average photon number. The application of loss transforms one coherent state to another with the mean photon number multiplied by the efficiency $\eta$. It is therefore impossible to calibrate the efficiency of a detector using coherent states without the use of a reference detector \cite{NAPHYS09}. We do not, therefore, account for the effects of loss in this measurement.  Cross-talk, on the other hand, changes the form of a coherent state distribution and allows us to accurately determine the cross-talk probability by measuring coherent states. To do this, we first take a measurement of the dark counts by blocking the laser in our setup. Denoting the measured single avalanche dark-count probability by $\varepsilon_{\mbox{\fontsize{5}{8}\selectfont ${D}$}}'$ it
can be shown that the dark count rate deducting the effect of cross-talk is
\begin{equation}
\varepsilon_{\mbox{\fontsize{5}{8}\selectfont ${D}$}}=\frac{\varepsilon_{\mbox{\fontsize{5}{8}\selectfont ${D}$}}'}{1-\varepsilon_{\mbox{\fontsize{5}{8}\selectfont ${XT}$}}} \, .\label{eq:dark-count}
\end{equation}
Next, we take a measurement of the laser pulses. We measured $10^6$ pulses in $\sim 10$ sec. Following the
notation of the previous section, we denote the measured avalanche number
probabilities by $\mathbf{p}'$ and the photon number probabilities
of the incoming state by $\mathbf{p}$. Using Eq. (\ref{eq:pt2p}) we
find the following relation between the first two elements of these
vectors,
\begin{subequations}
\begin{eqnarray}
p_1' &=& \Big( p_1 \left( 1-\varepsilon_{\mbox{\fontsize{5}{8}\selectfont ${D}$}} \right) +
p_0 \, \varepsilon_{\mbox{\fontsize{5}{8}\selectfont ${D}$}} \Big) \Big( 1-\varepsilon_{\mbox{\fontsize{5}{8}\selectfont ${XT}$}} \Big)
\label{eq:first}\\
p_0' &=& p_0 \left( 1-\varepsilon_{\mbox{\fontsize{5}{8}\selectfont ${D}$}} \right) .\label{eq:second}
\end{eqnarray}
\end{subequations}
We assume that $p_0=\exp(-\langle n \rangle)$ and $p_1=\exp(-\langle n \rangle) \times \langle n \rangle$, as expected from a coherent state. As
a result, we find that Eqs. (\ref{eq:first},\ref{eq:second}) form a set of two equations with two unknowns, $\langle n \rangle$ and
$\varepsilon_{\mbox{\fontsize{5}{8}\selectfont ${XT}$}}$ and can be solved to determine both of them. Using this method we calculated $\varepsilon_{\mbox{\fontsize{5}{8}\selectfont ${XT}$}}$ for a number of different
values of  $\langle n \rangle$ in the range $0.5-3.5$. As shown in Table \ref{tab:hresult}, we obtained $\varepsilon_{\mbox{\fontsize{5}{8}\selectfont ${XT}$}}=0.0975\pm0.0015$. This method for obtaining $\varepsilon_{\mbox{\fontsize{5}{8}\selectfont ${XT}$}}$ is more accurate than using only two photon dark-count rate as a measure of cross-talk
\cite{IEEENS07}.
\begin{table}[h]
\centering     
\begin{tabular}{c cccc}  
\hline
$\langle n \rangle$ & 0.86 & 1.66 &  2.30 & 3.13  \\  
\hline
$\varepsilon_{\mbox{\fontsize{5}{8}\selectfont ${XT}$}}$ & \quad 0.0960 \quad & \quad 0.0985 \quad & \quad 0.0965 \quad & \quad 0.0965 \quad\\
\hline
\end{tabular} \label{tab:hresult}
\caption{Measured cross-talk probability, obtained for coherent states with different mean photon numbers.}
\end{table}

\begin{figure}[h!]
\includegraphics{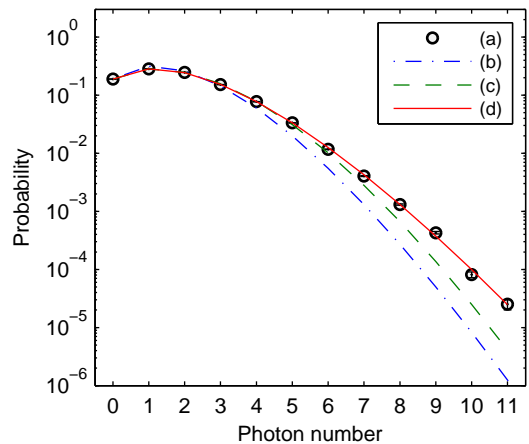}
\caption{\label{fig:pulseheight2} Measurement of a coherent state. (a) The raw avalanche statistics. (b) An uncorrected coherent state with $\langle n \rangle=1.66$ (c) The data in (b) multiplied by the matrix $M_{\mbox{\fontsize{5}{8}\selectfont ${XT}$}} \cdot M_{\mbox{\fontsize{5}{8}\selectfont ${D}$}}$ with first order cross-talk
compensation only. (d) The same as (c) but with full cross-talk compensation.}
\end{figure}

 The data measured for a coherent state with  $\langle n \rangle=1.66$ together with the compensated theoretical distribution is shown in Fig. \ref{fig:pulseheight2}. There is very close agreement between the experimental data and the theory over more than four and a half decades of probability values. The theoretical distribution with compensation for only first order cross-talk (i.e. up to one cross-talk event per pulse) is shown to illustrate that high order cross-talk compensation is necessary for obtaining an accurate fit at large photon numbers.

\section{\label{sec:quantummetrology}Quantum States: Tomography and Conditional Preparation}
\begin{figure}[]
\includegraphics{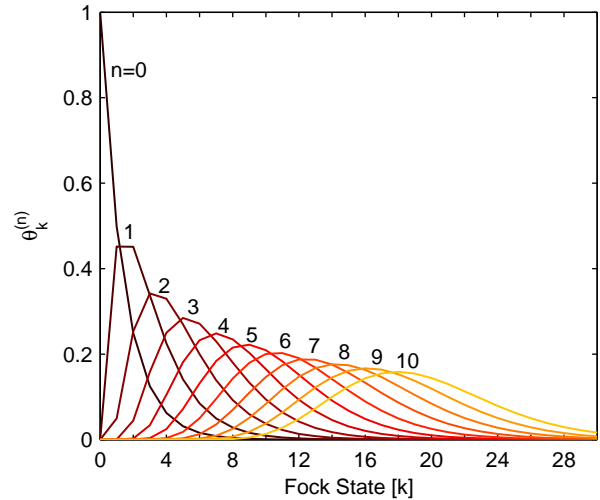}
\caption{\label{fig:povm} Elements of the MPPC's POVM, based on the model developed in Section \ref{sec:theoryanddataanalysis}. The different
curves correspond to events with different numbers of clicks. Results where calculated assuming a $50\%$ efficiency.}
\end{figure}

The complete characterization of quantum states of is known as quantum state tomography. Typically, when characterizing a state of light, it is assumed that there is an unlimited amount of copies at our disposal. A large number of measurement results are used to determine the most likely input state. In the following we use the term \textit{click} when referring to the discrete output of a general photon number resolving detector. An ideal detector would give exactly $n$ clicks when illuminated by an $n$ photon Fock-state. Unfortunately, existing detectors have various properties which obscure the one-to-one correspondence between the number of clicks and the number of impinging photons. The main properties that determine a given detector's applicability for quantum tomography are single photon detection efficiency and dark count rate. For time or space multiplexed detectors, the number of different detection modes is very significant. Since each mode can detect at most one photon, a large number of modes ensures that two photons will almost never reach the same mode. The large number of pixels in the MPPC's array make it very appealing in this respect. The existence of cross-talk, which is unique to multi-pixel detectors, must also be accounted for. The complete characterization of a quantum detector is given by its positive operator value measures (POVM) \cite{NAPHYS09}. Given an input density matrix
$\rho$, the probability $p_{n,\rho}$ of obtaining a detection outcome $n$ is
\begin{equation}
p_{n,\rho}=tr\left[\mathbf{\rho}\mathbf{\pi}_n\right],
\end{equation}
where $\left\{\mathbf{\pi}_n \right\}$ is the POVM. For a phase independent detector, $\pi_n$ is diagonal in the Fock state basis and has the
general form,
\begin{equation}
\displaystyle
\mathbf{\pi}_n=\sum_{k=0}^{\infty}\theta_k^{(n)}|k\rangle \langle
k|.
\end{equation}
Here, $\theta_k^{(n)}$ is the probability to obtain $n$ clicks when the detector is illuminated by a $k$ photon Fock state. It was shown in Eq.(\ref{eq:pt2p}) that the transformation between the illuminating state and the click statistics is given by,
 \begin{equation}
\mathbf{M}_{TOT}=\mathbf{M}_{XT}\cdot\mathbf{M}_{D}\cdot\mathbf{M}_{L} \, . \label{eq:mtot}
\end{equation}
The matrix $\mathbf{M}_{TOT}$ contains all the information needed for constructing the MPPC's POVM. The connection between the two being given
by,
 \begin{equation}
\left[ M_{TOT} \right]_{k,n} =\theta_k^{(n)}. \label{eq:mtot}
\end{equation}
The elements of the POVM with an assumed efficiency $\eta$ are shown in Fig. \ref{fig:povm}. The normalization is such that $\sum_n \theta_k^{(n)}=1$. Note that in the absence of cross-talk, the peak of the $n$ click curve occurs at $i\simeq n/ \eta$. Due to cross-talk, the peak is shifted slightly to the left i.e. to lower Fock states since an $n$ click event may originate from a Fock state with less than $n$ photons. Due to the large number of spatial modes, the POVM's shape is determined mainly by the detection efficiency. For time-multiplexed detectors the limited number of modes manifests itself in a broadening of the POVM curves and a significant shift towards higher photon numbers \cite{NAPHYS09}. We conclude that multi-pixel detectors are applicable to quantum state tomography. The existence of cross-talk doesn't pose a significant limitation and the large number of pixels enhances the ability to reconstruct high photon number states.

\begin{figure}[]
\includegraphics{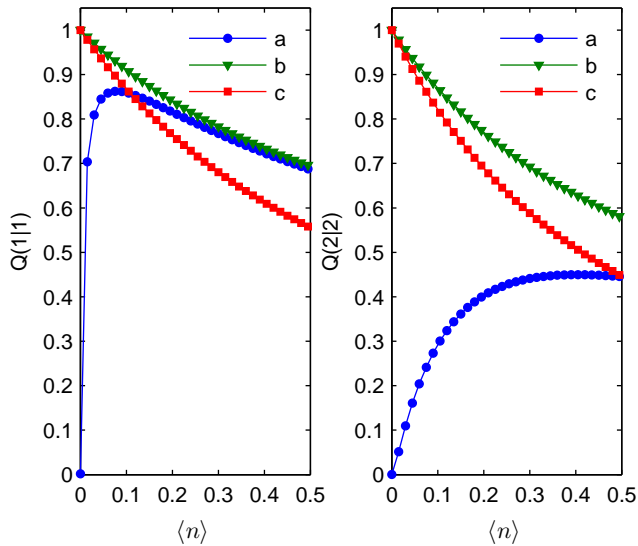}
\caption{\label{fig:spp} Fidelity of the prepared state for a single photon Fock state (left pane) and a two photon Fock state (right pane), as a function of the average number of photons in either of the modes, calculated results. The post-selection scheme is described in the text. The fidelity is plotted for three different heralding detectors: (a) the complete MPPC  model, (b) the MPPC model with loss but without cross-talk and dark-counts , (c) a single APD (left pane) two APDs' and a beamsplitter (right pane).  }
\end{figure}

In entanglement-based quantum state preparation a measurement on one mode of an entangled two mode state predicts the existence of a desired state in the other mode \cite{JMODOP04a,PRA01}. Quantum state preparation is most commonly realized using two-mode spontaneous parametric down-conversion (SPDC). The wave-function of the two mode state before detection is given by \cite{IntroQuantOPt}
\begin{gather}
| \psi \rangle = \sum_{n=0}^{\infty}C_n |n\rangle_{a}|n\rangle_{b} \notag
\\
C_n=\frac{1}{\cosh r}  (-1)^n e^{\imath n \theta} \tanh^nr \notag
\\
\langle n \rangle_a=\langle n \rangle_b=\sinh^2r
\label{eq:spdca}.
\end{gather}
The detection of $n$ photons in mode $a$ with an \emph{ideal} detector indicates the existence of $n$ photons in mode $b$. These are so-called heralded $n$ photon Fock states. For a realistic detector one defines the fidelity $Q(k|k)$ as the probability that given $k$ clicks in mode $a$ there are actually $k$ photons in mode $b$. In terms of the POVM elements $Q(k|k)$
is written as
\begin{gather}
Q(k|k)=\frac{\theta^{(k)}_k \cdot |C_k|^2}{\sum_{i=0}^{\infty} \theta^{(k)}_i \cdot |C_i|^2} \notag \label{eq:spdcb}.
\end{gather}
To illustrate the effect of dark-count and cross-talk on the preparation fidelity, $Q(k|k)$ is plotted in Fig \ref{fig:spp} for $k=1,2$.
For each value of $k$ we plot the fidelity for both the complete MPPC  model and the MPPC without cross-talk and dark-counts. This can be done by omitting the dark count and the cross-talk matrices from $\mathbf{M}_{TOT}$. In the absence of a photon number resolving detector one could detect a single photon using a single APD and two photons using two APDs' and a $50/50$ beamsplitter \cite{PRL06}. The fidelity of state preparation with heralding performed using these basic solutions (assuming zero dark count and $50\%$ efficiency) is plotted for comparison. For $k=1$ it can be seen that approaching small values of $\langle n \rangle$ the fidelity starts to curve downwards and finally reaches zero at values when $\langle n \rangle$ is comparable to the dark count rate. This reduction of fidelity occurs due to dark-counts and has nothing to do with cross-talk. Nevertheless, for mean photon numbers larger than $0.1$ the MPPC obtains better fidelity than a single APD despite the dark-counts. For $k=2$ the situation is even less favorable. The reason being that single photon events are mistaken for two  photon events due to cross-talk. We conclude that cross-talk significantly impedes the use of multi-pixel detectors for quantum state preparation at least in realistic scenario described above. The elimination of cross-talk as well as the reduction of dark counts would be desirable for this application.

\section{Conclusion}
We have studied the use of multi-pixel detectors in the context of quantum optics using the Hamamatsu MPPC as a prototype. This device is shown to have excellent photon number resolution.  Using triggered data acquisition and post selection we achieved a room temperature dark-count rate of $\sim 2\times 10^{-3}$ per pulse which may be further reduced by cooling. In addition, we experimentally measured the cross-talk probability with high precision using coherent states. A complete model for photon number reconstruction was developed which includes the effect of high order cross-talk. Due to the large number of pixels, such detectors are particularly suitable for measurement of states with high photon numbers, making them attractive candidates for quantum state tomography. On the other hand, the fidelity of quantum states prepared using a standard entanglement-based protocol is found to be low due to the effects of cross-talk and dark counts. The full potential of these detectors will be realizable when cross-talk between pixels is eliminated.

\begin{acknowledgments}
Itai Afek gratefully acknowledges the support of the Ilan Ramon Fellowship.
Financial support of this research by the German Israeli
foundation (GIF) is gratefully acknowledged.
\end{acknowledgments}


\end{document}